\documentclass[12pt,reqno]{amsart}
\usepackage{amsmath}
\usepackage{amsgen,amstext,amsbsy,amsopn,amsthm,amssymb,amsxtra}
\usepackage{graphicx}

\usepackage{pdfsync}
\usepackage{graphicx}

\theoremstyle{definition}

\newcommand{\half}{\mbox{$\frac{1}{2}$}}
\newcommand{\beq}{\begin{equation}}
\newcommand{\eeq}{\end{equation}}

\newcommand{\version}{July 13, 2014}

\theoremstyle{definition}

\usepackage{color}

\newcommand{\E}{\mathcal{E}}
\renewcommand\epsilon\varepsilon

\theoremstyle{definition}

\newcommand{\R}{\mathbb{R}}

\newcommand{\eps}{\epsilon}
\renewcommand{\epsilon}{\varepsilon}

\renewcommand{\phi}{\varphi}

\newcommand{\Z}{\mathbb{Z}}

\begin{document}

\title[Superfluid behavior in a random potential]{{\bf 
 Superfluid behavior of a Bose-Einstein condensate \\ in a random potential 
}}
\author[M. K\"onenberg]{Martin K\"onenberg}
\address{M. K\"onenberg, Department of Mathematics and Statistics, Memorial University \\ \it St. Johns, NL, Canada}
\email{mkonenberg@mun.ca}
\author[T. Moser]{Thomas Moser}
\address{T. Moser, Fakult\"at f\"ur Physik, Universit{\"a}t Wien	\\ \it Boltzmanngasse 5, 1090 Vienna, Austria}
\email{thomas.moser@univie.ac.at}
\author[R. Seiringer]{Robert Seiringer}
\address{R. Seiringer, Institute of Science and Technology Austria (IST Austria) 
\\ \it Am Campus 1, 3400 Klosterneuburg, Austria}
\email{robert.seiringer@ist.ac.at}
\author[J. Yngvason]{Jakob Yngvason}
\address{J. Yngvason, Fakult\"at f\"ur Physik, Universit{\"a}t Wien	\\ \it Boltzmanngasse 5, 1090 Vienna, Austria}
\email{jakob.yngvason@univie.ac.at}

\date{
\version}
\newcommand\const{{\rm const\,}}
\begin{abstract}
We investigate the relation between Bose-Einstein condensation (BEC) and superfluidity in the ground state of a one-dimensional model of interacting Bosons in a strong random potential. We prove rigorously that in a certain parameter regime the superfluid fraction can be arbitrarily small while complete BEC prevails. In another regime there is both complete BEC and complete superfluidity, despite the strong disorder. 
\end{abstract}
\maketitle
\section{Introduction}
One of the intruiging issues in the theory of superfluidity is the question of its relation to Bose-Einstein condensation (BEC). A precise formulation of this question requires precise definitions of the concepts. In the case of BEC, the universally accepted definition is in terms of the macroscopic occupation of some one particle state, measured by the largest eigenvalue of the one particle density matrix of the many body state \cite{PO}. In the case of superfluity, on the other hand, the definition is not so clear cut. As emphasized by Leggett \cite{Leggett} one must distinguish between dynamical aspects like frictionless flow at a finite speed, and the response of the system to an infinitesimally small imposed velocity field, e.g. through slow rotation of a container. The latter is  easier to handle mathematically and in the sequel we shall use the customary definition of the superfluid fraction as the second derivative with respect to the velocity at zero of the energy per particle \cite{HoM}. This definition can equivalently be formulated in terms of twisted boundary conditions. 

For liquid helium 4 there is experimental  and numerical evidence for almost complete superfluity near absolute zero while the BEC fraction is less than 10 \% \cite{Sokol}. Also, a one-dimensional hard-core Bose gas is an example of a system that is superfluid in the ground state but where BEC is absent. In general it has been argued, see. e.g. \cite{H}, that neither condition is necessary for the other, and that disorder may destroy superfluidity while BEC prevails \cite{ABCG, KT, yukalovgraham, YY}. A mathematical investigation of this question, however, is hampered by the fact that a rigorous proof of BEC in a system of interacting Bosons is a notoriously difficult problem that has only been solved in a few special cases. One such case is the proof of both complete BEC and complete superfluidity in the ground state of a dilute  Bose gas in a smooth trapping potential in the Gross Pitaevskii limit \cite{LS02, LSYsuper}.

In recent years the interplay between interactions and disorder in many body systems has been studied in many works,  both theoretically and experimentally. It is not the intention here to give a review of the subject but we mention the references \cite{gimperlein}--\cite{BW} as a representative sample. In \cite{SYZ} (see also \cite{SYZ2}) a one dimensional model of an interacting Bose gas was studied and it was shown that complete BEC in the ground state may survive a strong random potential in an appropriate limit. On the other hand, the random potential may have drastic effects on the wave function of the condensate and this can be expected to influence the superfluid behavior. In the present paper we analyze the density distribution of the condensate in this model in some detail and its implications for superfluity.  Our main result  is a rigorous proof that in a certain parameter regime the superfluid fraction can be arbitrarily small although complete BEC prevails,  while in another regime there is both complete BEC and complete superfluidity.

The model we consider is the Lieb-Liniger model \cite{LL} of bosons with
contact interaction on the unit interval but with an additional
{external random potential} $V_\omega$. The Hamiltonian on the Hilbert
space $L^2([0,1], dz)^{\otimes_{\rm symm}N}$ of square integrable symmetric wave functions of $(z_1,\dots,z_N)$ with $0\leq z_i\leq 1$ is
\begin{equation} \label{model}
  H=\sum_{i=1}^N\left(-\partial_{z_i}^2+V_\omega(z_i)\right)+\frac{{\gamma}}
  N\sum_{i<j}\delta(z_i-z_j)
\end{equation}
where $\gamma\geq 0$ and we shall take periodic boundary conditions for the kinetic energy operator. The random
potential is taken to be
\begin{equation}\label{randompot}
  V_\omega(z)=\sigma
  \sum_j\delta(z-z_j^\omega)
\end{equation}
with $\sigma \geq 0$ independent of the random sample $\omega$ while
the \emph{obstacles} $\{z_j^\omega\}$ are \emph{Poisson distributed}
with density $\nu\gg 1$, i.e., their mean distance is $\nu^{-1}$. 
The Hamiltonian \eqref{model} can be defined rigorously via the quadratic form on the Sobolev space  $H^1([0,1]^{\otimes N})$  given by the expression on the right hand side of \eqref{model}, noting that functions in the Sobolev space can be restricted to hyperplanes of codimension 1. (The Sobolev space  $H^1([0,1]$ consists of functions on $[0,1]$ that together with their first derivative are square integrable.) 

Since our model is formulated in the fixed interval $[0,1]$ the particle density $\rho$ tends to infinity as $N\to\infty$. Equivalently, we could have considered the model in an interval $[-L/2, L/2]$ and taking $N$ and $L\to\infty$ with $\rho=N/L$ fixed, as done for instance in \cite{BW}. As explained in  \cite{SYZ2}, Sect.\ 4.4,  the two viewpoints are connected by simple scaling. 
For the purpose of the present investigation we find it convenient to stick to the model in the unit interval.

Besides the particle number $N$ the model \eqref{model} has three parameters: $\nu$, $\gamma$ and $\sigma$. The limiting case $\sigma=\infty$ amounts to requiring
the wave function to vanish at the positions of the obstacles $z^\omega_j$. In \cite{SYZ} it is shown that as $N\to\infty$ for fixed values of the parameters Bose-Einstein condensation takes place in the ground state. The ground state energy and the wave function of the condensate is described by a  Gross Pitaevskii (GP) energy functional (see below, Eq. \eqref{gp}). In \cite{SYZ} it is furthermore proved that the corresponding energy becomes deterministic, i.e., independent of $\omega$ in probability, if the parameters satisfy the conditions
\begin{equation}\label{conditions}
    \nu\gg 1\,, \quad \gamma\gg \frac{\nu}{\left(\ln \nu\right)^2}\,, \quad 
    \sigma \gg \frac{ \nu}{1+ \ln \left( 1+ \nu^2/\gamma\right) }\,.
  \end{equation} 
In \cite{SYZ} it is explained why these conditions are also necessary in order to obtain a deterministic energy and they will be presupposed in all statements about the model in the sequel.
  
Our new findings  about the model can be summarized as follows.\smallskip

\noindent{\bf Main Results:}\smallskip

{\it \begin{itemize}
\item If $\gamma\lesssim \nu^2$ the superfluid fraction is arbitrarily small, i.e., it goes to zero in the limit \eqref{conditions}.\medskip
\item The same holds for $ \nu^2\ll \gamma\ll \nu^4$ provided $\sigma\gg(\gamma/\nu^2)^2\gamma^{1/2}$.\medskip
\item If $\gamma\gg (\sigma \nu)^2$ there is complete superfluidity, i.e., the superfluid fraction tends to 1.
\end{itemize}}\medskip

The estimates that lead to these assertions are contained in Eqs. \eqref{sobolev}, \eqref{3.10} and  \eqref{sfest} below. Figure 1 illustrates the parameter regions with and without superfluidity.  (Note that we are concerned with asymptotic parameter regimes and the boundaries of the colored areas are not meant to indicate sharp transitions.)
\begin{figure}[htf]
\center
\fbox{\includegraphics[width=8cm]{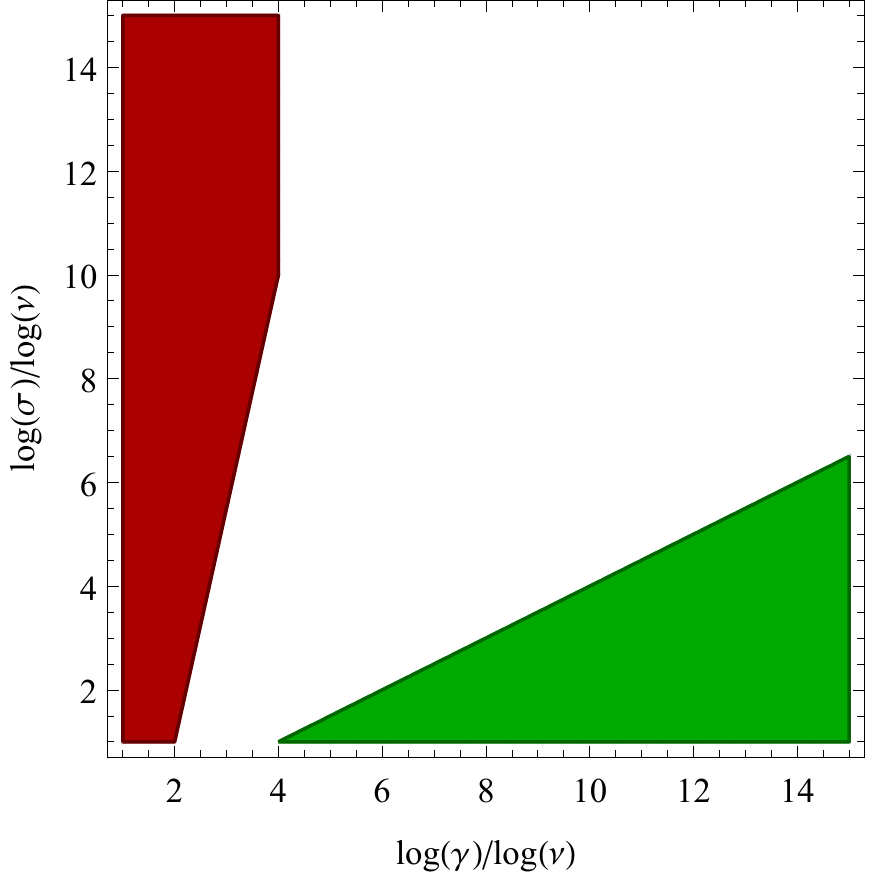}}
\caption{Red: Absence of superfluidity. Green: Complete superfluidity.}
\label{}
\end{figure}
\medskip

We now describe briefly the organization of the paper. In the next Section 2 we first recall from \cite{SYZ} the description of the ground state properties of the Hamiltonian \eqref{model}, in particular BEC, in terms of a Gross Pitaevskii functional. For this it is not necessary to assume the special potential \eqref{randompot}, and we can state the results for an arbitrary nonnegative potential $V$. The same holds in Section 2.2 where we show that superfluity in the ground state of many body Hamilonian is, in the large $N$ limit, equivalent to superfluity described in terms of the GP theory. In Section 3 we shall derive a  {\it closed formula for the superfluid fraction} $\rho^{\rm sf}$:
\begin{equation}\label{defsf}
\rho^{\rm sf} = \left( \int_0^1 |\psi_0(z)|^{-2} dz \right)^{-1}
\end{equation}
where $\psi_0$ is the minimizer of the GP energy functional. 

A further general result (for an arbitrary nonnegative potential $V$) that we prove in Section 4 is an estimate for the deviation of the density from 1 in the sup norm $\Vert\cdot\Vert_\infty$:
\beq\label{sobolev}
 \frac { \| |\psi_0|^2 - 1 \|_\infty^2 } { \sqrt{ 1 + \| |\psi_0|^2 - 1 \|_\infty}} \leq \frac {2^{3/2}} {\sqrt{\gamma}} \int_0^1 V\,.
\eeq
When applied to $V=V_\omega$ this leads immediately to the  sufficient criterion $\gamma\gg (\sigma \nu)^2$ for complete superfluidity.

The absence of superfluity in the random potential for weak interactions and/or high density of scatterers is derived in Section 5.

\section{ BEC and superfluidity in the GP limit}
\subsection{BEC}An important fact about the Hamiltonian \eqref{model} that was proved in \cite{SYZ} is Bose-Einstein condensation in the ground state in the limit when $N\to\infty$ and $\gamma$ is fixed (GP limit), or does not grow too fast with $N$. This holds in fact also if $V_\omega$ is replaced by an an arbitrary positive potential $V$. The wave function of the condensate (eigenfunction to the highest eigenvalue of the one particle density matrix) is the minimizer $\psi_0$ of the Gross Pitaevskii (GP) energy functional
\beq\label{gp}
\mathcal{E}^{\rm GP}[\psi] = \int_0^1 \left( |\psi'(z)|^2 + V(z) |\psi(z)|^2 + \frac \gamma 2 |\psi(z)|^4 \right) {\mathrm d} z
\eeq
with the normalization $\int_0^1|\psi|^2=1$. The minimizer $\psi_0$ is also the ground state of the mean field Hamiltonan
\beq\label{meanfieldham}
h=-\partial_{z}^2+V(z)+\gamma |\psi_0|^2-\frac\gamma 2\int_0^1 |\psi_0|^4
\eeq
with eigenvalue $e_0=\mathcal{E}^{\rm GP}[\psi_0]$. The average occupation of the one particle state $\psi_0$ in the many-body ground state $\Psi_0$ of $H$ is $N_0=\langle \Psi_0, a^\dagger(\psi_0)a(\psi_0)\Psi_0\rangle$ with $a^\dagger(\psi_0)$ and $a(\psi_0)$ the creation and annihilation operators for $\psi_0$. Bose-Einstein condensation is expressed through the estimate

\beq\label{depletion} \left(1-\frac {N_0}N\right)\leq ({\rm const.}) \frac {e_0}{e_1-e_0}
N^{-1/3}\min\{\gamma^{1/2},\gamma\}
\eeq
where $e_1$ is the second lowest eigenvalue of the mean field Hamiltonian \eqref{meanfieldham}. Moreover,  the ground state energy per particle of $H$  converges to the GP energy $e_0$, see Eqs. \eqref{upperbd} and \eqref{lowerbd} below.
\subsection{Superfluidity}

To discuss superfluidity we modify the kinetic term of the Hamiltonian, replacing $\mathrm i\partial$ by $\mathrm i\partial+v$ with a velocity $v\in\mathbb R$. We thus consider 
\beq
H_v=\sum_{j=1}^N\left\{({\mathrm i}\partial_{z_j}
+v)^2+V(z_j)\right\}+\frac \gamma N\sum_{i<j}\delta(z_i-z_j)
\eeq
on $L^2([0,1],{\mathrm d}z)^{\otimes^N_{\rm symm}}$ with periodic boundary conditions. Let $E_0^{\rm QM}(v)$ denote its ground state energy and 
let $e_0(v)$ denote the corresponding ground state energy of the modified GP functional
\beq\label{modGP}
\mathcal{E}^{\rm GP}_v[\psi] = \int_0^1 \left( |{\mathrm i}\psi'(z)+ v\psi(z)|^2 + V(z) |\phi(z)|^2 + \frac \gamma 2 |\psi(z)|^4 \right) {\mathrm d} z.
\eeq
For small enough $v$, $\mathcal{E}_v$ has a unique minimizer, denoted by $\psi_v$, and $e_0(v)$ is equal to the ground state energy of the mean field Hamiltonian
\beq
h_v=({\mathrm i}\partial_{z}+v)^2+V(z)+\gamma |\psi_v(z)|^2-\frac\gamma 2\int_0^1 |\psi_v|^4.
\eeq
Taking $\psi_v^{\otimes N}$ as trial function for the Hamiltonian $H_v$
we obtain
 \beq\label{upperbd}
 E_0^{\rm QM}(v)/N\leq e_0(v).
 \eeq
 For the lower bound we write in the same way as Eq. (7) in \cite{SYZ}
 \begin{multline}
H_v=\sum_{j=1}^N\left\{\left(1-\frac{N-1}{2N} \eps\right)({\mathrm i}\partial_{z_j}
+v)^2+V(z_j)\right\}\\+\frac 1N\sum_{i<j}\left[\frac \eps 2(({\mathrm i}\partial_{z_i}
+v)^2+({\mathrm i}\partial_{z_j}
+v)^2)+\gamma\delta(z_i-z_j)\right].
\end{multline}
We may now use the diamagnetic inequality (\cite{LL}, p.\ 193)  to bound an expectation value of the second term with respect to any wave function $\Psi$ from below by the expectation value of 
\beq
\frac 1N\sum_{i<j}\left[\frac \eps 2(-\partial_{z_i}^2
-\partial_{z_j}^2)
+\gamma\delta(z_i-z_j)\right]
\eeq
with respect to $|\Psi|$. Proceeding exactly as in \cite{SYZ}, Eqs. (12)-(17), we can thus  bound $H_v$ from below in terms of the mean field Hamiltonian with controlled errors terms, arriving at the lower bound for the ground state energy
\beq\label{lowerbd}
E_0^{\rm QM}(v)/N\geq e_0(v)(1-(\const)N^{-1/3}\min\{\gamma^{1/2},\gamma\}).
\eeq 
We conclude that in the GP limit the superfluid fraction
\beq\label{sfracQM}
\rho^{\rm sf}=\lim_{v\to 0}\frac 1{v^2}\lim_{N\to\infty}\frac 1N(E_0^{\rm QM}(v)-E_0^{\rm QM}(0))
\eeq
is the same as the corresponding quantity derived from the GP energy, i.e., 
\beq\label{sfracgp}
\rho^{\rm sf}=\lim_{v\to 0}\frac 1{v^2}(e_0(v)-e_0(0)).
\eeq
Note that in \eqref{sfracQM} the order in which the limits are taken is important in general. Using the GP minimizer for $v=0$ as a trial state for \eqref{modGP} we see that $\rho^{\rm sf}\leq 1$.
Note also that the error term in \eqref{lowerbd} is independent of $V$ and uniformly small in $\gamma$ for
$\gamma\ll N^{2/3}$.

\section {Proof of Eq.\ \eqref{defsf}}

In this section we shall prove the formula \eqref{defsf} for the superfluid density. We start with the variational equation for $\psi_v$ which is
\beq
(i \partial_z + v)^2 \psi_v(z) + V(z) \psi_v(z) + \gamma |\psi_v(z)|^2 \psi_v(z) = \mu \psi_v(z).
\eeq
We multiply this by $\bar\psi_v$ and take the imaginary part, to obtain
\beq
\partial_z \left( v |\psi_v(z)|^2 - \Im [ \bar\psi_v(z) d\psi_v(z)/dz] \right) = 0
\eeq
hence there exists a constant $C\in \R$ such that 
\beq
 \Im [ \bar\psi_v(z) d\psi_v(z)/dz] = v |\psi_v(z)|^2 - C.
 \eeq
 Since 
 \begin{equation}\label{primeE}
de_0(v)/dv = 2 v - 2  \int_0^1 \Im [\bar \psi_v(z) d\psi_v(z)/dz] dz
\end{equation}
 we actually see that $C=\half de_0(v)/dv$. 
 For small $v$, $\psi_v$ has no zeroes, hence we can divide by $|\psi_v(z)|^2$ and obtain
 \beq
 S'(z): = \frac{  \Im [ \bar\psi_v(z) d\psi_v(z)/dz]  }{ |\psi_v(z)|^2}  = v - \frac C {|\psi_v(z)|^2}.
 \eeq
 Since $S'$ is, in fact, the derivative of the phase of $\psi_v$, i.e, $\psi_v(z) = |\psi_v(z)| e^{iS(z)}$, 
 we have, for a  system with periodic boundary conditions, 
 \beq
 \int_0^1 S'(z) dz = 2\pi n 
 \eeq
 for $n\in \Z$. For small enough $v$, one has $n=0$, and hence 
 \beq
 v = C \int_0^1 |\psi_v(z)|^{-2} dz. 
 \eeq

We plug this into (\ref{primeE}), to obtain
\beq
e_0'(v) = 2 C = 2 v \left(  \int_0^1 |\psi_v(z)|^{-2} dz  \right)^{-1}.
\eeq
With 
\beq
\rho^{\rm sf} = \lim_{v\to 0} \frac{ e_0'(v)}{2v}
\eeq
this leads to the formula (\ref{defsf}). 

\section {Proof of Eq. \eqref{sobolev}}

We now derive the bound (5) which quantifies the deviation of the GP minimizer from a constant in terms of the average value of the random potential and the interaction strength.

Functions $f$ in the Sobolev space $H^1([0,1])$ are continuous, and hence  $\int_0^1 f = 0$ implies that $f(z) = 0$ for some $z\in [0,1]$. For such $f$, we have 
\beq
f^2(x) =2  \int_z^x f'(y) f(y) dy 
\eeq
and hence
\beq
\|f\|_\infty^2 \leq 2 \|f'\|_2 \|f\|_2\,,
\eeq
where $\Vert\cdot\Vert_\infty$ is the sup norm and  $\Vert\cdot\Vert_2$ the $L^2$-norm. We apply this to $f(x) = |\psi(x)|^2 - 1$ for an $L^2$-normalized function $\psi$. This gives
\beq
\| |\psi|^2 - 1 \|_\infty^2 \leq 4 \| \psi' \psi \|_2 \| |\psi|^2 -1 \|_2 \leq 4 \|\psi'\|_2 \|\psi\|_\infty \| |\psi|^2 -1\|_2\,.
\eeq
We further bound $\|\psi\|_\infty^2  \leq 1 + \| |\psi|^2 -1 \|_\infty$ and hence find
\beq
 \|\psi'\|_2  \| |\psi|^2 -1\|_2 \geq \frac 14 \frac { \| |\psi|^2 - 1 \|_\infty^2 } { \sqrt{ 1 + \| |\psi|^2 - 1 \|_\infty}}\,.
 \eeq
 In particular, for $\gamma>0$, 
\begin{align}
\| \psi'\|_2^2 + \frac \gamma 2 \int_0^1 |\psi|^4 & = \| \psi'\|_2^2 + \frac \gamma 2 \| |\psi|^2 -1 \|_2^2 + \frac \gamma 2 \nonumber\\ & \geq \frac \gamma 2 + {\sqrt{2\gamma}}  \|\psi'\|_2  \| |\psi|^2 -1\|_2 \nonumber\\ & \geq \frac \gamma 2 +  \frac{\sqrt{\gamma}}{2^{3/2}} \frac { \| |\psi|^2 - 1 \|_\infty^2 } { \sqrt{ 1 + \| |\psi|^2 - 1 \|_\infty}}\,.
\end{align}
For $V\geq 0$ the GP minimizer $\psi_0$ satisfies (take $\psi\equiv 1$ as a trial function)
\beq
\| \psi_0'\|_2^2 + \frac \gamma 2 \int_0^1 |\psi_0|^4\leq \| \psi_0'\|_2^2 + \int_0^1 V(z) |\psi_0(z)|^2 dz + \frac \gamma 2 \int_0^1 |\psi_0|^4\leq \int_0^1 V+\frac \gamma 2, \eeq
so
\beq
 \frac { \| |\psi_0|^2 - 1 \|_\infty^2 } { \sqrt{ 1 + \| |\psi_0|^2 - 1 \|_\infty}} \leq \frac {2^{3/2}} {\sqrt{\gamma}} \int_0^1 V\,.
\eeq
Since  $\int_0^1 V_\omega$ is close to $\nu\sigma$ with high probability, in the sense that the ratio converges to 1 in probability, we see that $|\psi_0|^2$ and hence also $|\psi_0|^{-2}$ converges uniformly to 1 if $\gamma\gg(\sigma \nu)^2$ as the parameters tend to $\infty$. Thus the superfluid fraction is equal to 1 by Eq. \eqref{defsf}.

\section{Absence of superfluidity }

If $\mathcal I$ is any (measurable) subset of $[0,1]$ with length $|\mathcal I|$ it follows from Eq.\ (4) and the Cauchy Schwarz inequality that 
\beq\label{superfbound} \rho^{\rm sf}\leq \frac{\int_{\mathcal I}|\psi_0(z)|^2 dz}{|\mathcal I|^2}.
\eeq
To prove that superfluidity is small we have therefore to identify subsets such that $\int_{\mathcal I}|\psi_0(z)|^2 dz$ is small, while $|\mathcal I|$ is not too small. 

The random points $z_j^\omega$ split the intervall $[0,1]$ into subintervals
$\mathcal I_j=[z_j^\omega, z_{j+1}^\omega]$ of various lengths $\ell_j=z_{i+1}^\omega-z_{j}^\omega$. The lengths are independent random variables\footnote{Strictly speaking, because of the fixed endpoints 0 and 1, the interval lengths are not  quite independent, but since the number of intervals is very large this does not affect the estimates.}
with identical probability distribution
\beq dP_\nu(\ell)=\nu e^{-\nu\ell}\, d\ell.
\eeq
We anticipate that intervals of small lengths have small occupation and shall therefore take
\beq \mathcal I=\bigcup_{j:\ell_j\leq\tilde \ell}\mathcal I_j
\eeq
with a suitably chosen $\tilde\ell$. The average length of $\mathcal I$ is
\beq \label{3.12} L=\nu\int_0^{\tilde\ell}\ell dP_\nu(\ell)=1-(1+(\nu\tilde\ell))e^{-\nu\tilde\ell}).\eeq
In particular it tends to 1 if and only if $\tilde\ell\gg \nu^{-1}$.

With the notation
\beq\label{gpmass}
 n_j^{\rm GP}=\int_{\mathcal I_j}|\psi_0(z)|^2dz\eeq
we define
\beq \label{3.5} N_{s,\omega}=\int_{\mathcal I}|\psi_0(z)|^2 dz=\sum_{\ell_j\leq \tilde \ell} n^{\rm GP}_j.\eeq
Note that $\psi_0$ and $n_j^{\rm GP}$ also depend on $\omega$ but we have suppressed this in the notation for simplicity.

Our estimate on $N_{s,\omega}$ is based on estimates on the GP energy that were derived in \cite{SYZ}. These involve some auxiliary quantities that we now recall.

\subsection{The energy between obstacles}

The energy in an interval where the obstacles are placed only at the endpoints is given by suitable rescaling of the energy functional
\begin{equation}
\E_{\kappa,\alpha}[\phi]= \int_0^1 dx \ \left( |\phi'(x)|^2 + \frac \kappa 2 |\phi(x)|^4\right) +
\frac \alpha 2 \left( |\phi(0)|^2 + |\phi(1)|^2 \right)  \label{GP-funct}
\end{equation}
with $\kappa\geq 0$ and $\alpha\geq 0$. Let $e(\kappa,\alpha)$ denote the auxiliary GP energy
\begin{equation}\label{GP-energ}
e(\kappa,\alpha) = \inf_{\|\phi\|_{2} =1} \E_{\kappa,\alpha}[\phi] \ .
\end{equation}
The corresponding energy for an interval of length $\ell$ with mass $\int_{\rm Interval}|\phi|^2=n$, coupling constant $\gamma$ and strength $\sigma$ of the obstacle potential is then, by scaling,
\beq\frac n{\ell^2}e(n\ell\gamma, \ell\sigma).
\eeq
We shall use the following bounds on $e(\kappa,\alpha)$ that were derived in \cite{SYZ}, Eqs. (32) and (41):
\begin{equation}\label{tos}
e(\kappa,\infty) \geq e(\kappa,\alpha) \geq e(\kappa,\infty) \left( 1 - K \alpha^{-1/2} \right) 
\end{equation} 
and
\begin{equation}\label{e0a}
e(\kappa,\alpha)\geq e(0,\alpha) \geq \frac {C\alpha}{1+\alpha} \,.
\end{equation}
with  constants $K$ and $C$ independent of $\kappa$ and $\alpha$.

\subsection{The interval density functional}

With $n(\ell)\geq 0$ a mass distribution on intervals of various lengths $\ell$ we define an ``interval density functional'', cf. \cite{SYZ}, Eq. (42), as
 \beq 
\mathcal E^{\rm IDF}[n(\cdot)]=\nu\int_0^\infty dP_\nu(\ell)\frac {n(\ell)}{\ell^2}e(n(\ell)\ell\gamma, \infty)
\eeq
with corresponding energy
\beq\label{minprob}
e^{\rm IDF}(\nu,\gamma)=\inf\left\{ \mathcal E^{\rm IDF}[n(\cdot)]\,:\, \nu \int_0^\infty dP_\nu(\ell)n(\ell)=1 \right\}.
\eeq
 This energy (denoted by $e_0(\gamma,\nu)$) is in \cite{SYZ}, Theorem 3.1,  proved to be the deterministic limit (in probability) of the GP energy under the conditions \eqref{conditions}. The minimization problem \eqref{minprob} is conveniently treated by introducing  a Lagrange multiplier $\mu$ for the normalization condition on $n(\ell)$. In \cite{SYZ}, Eqs. (45)--(47), it is shown that 
 \begin{equation}\label{rel-1}
\mu \sim \gamma\, f(\nu^2/\gamma),\end{equation}
where $f:\mathbb{R}_+\to \mathbb{R}_+$ denotes the function
\begin{equation}\label{def:f}
f(x) = \left\{ \begin{array}{ll} 1 & \text{for $x\leq 1$} \\ \frac x{ \left(1+ \ln x\right)^2} & \text{for $x\geq 1$.} \end{array} \right.
\end{equation}
Also $e^{\rm IDF}(\gamma,\nu) \sim \gamma\, f(\nu^2/\gamma)$. A further result derived in \cite{SYZ} is that the minimizing $n(\ell)$  of the interval density functional is nonzero if and only 
if $\mu\ell^2>\pi^2$. We can therefore expect that the mass \eqref{gpmass} is small in intervals $\mathcal I_j$ such that
$\ell_j\leq (\const.)/\sqrt \mu$ and we shall make use of this in the following considerations.

\subsection{Absence of superfluidity for $\gamma\lesssim \nu^2$}

The first step is to split the GP energy $e_\omega^{\rm GP}(\gamma,\nu,\sigma)$, which is the minimum energy of \eqref{gp} with $V_\omega$ in place of $V$, 
into  contributions from `large' and `small' intervals:
\beq\label{3.1}
e_\omega^{\rm GP}(\gamma,\nu,\sigma)\geq \sum_{\ell_j\geq \tilde \ell}\frac {n_j^{\rm GP}}{\ell_j^2}e(n_j^{\rm GP}\ell_j\gamma, \ell_j\sigma)+ \sum_{\ell_j< \tilde \ell}\frac {n_j^{\rm GP}}{\ell_j^2}e(n_j^{\rm GP}\ell_j\gamma, \ell_j\sigma)
\eeq
where
\beq\label{3.2} \tilde\ell=s/\sqrt \mu
\eeq
with a suitable $s$ to be chosen later, and (by \eqref{rel-1} and \eqref{def:f})
\beq\label{3.3}
\mu\sim \frac {\nu^2}{(1+\ln(\nu^2/\gamma))^2}.
\eeq
Note that, since $\sigma\gg \nu/(1+\ln(1+\nu^2/\gamma))$ we have
\beq\label{3.4}
\tilde\ell \sigma \gg1
\eeq
for $\gamma\lesssim \nu^2$.

We estimate the sum over the small intervals using Eqs. \eqref{e0a} and \eqref{3.2}:
\begin{multline}\label{3.6}
\sum_{\ell_j< \tilde \ell}\frac {n_j^{\rm GP}}{\ell_j^2}e(n_j^{\rm GP}\ell_j\gamma, \ell_j\sigma)
\geq \sum_{\ell_j< \tilde \ell}\frac {n_j^{\rm GP}}{\ell_j^2}e(0, \ell_j\sigma)
\geq \sum_{\ell_j< \tilde \ell}\frac {n_j^{\rm GP}}{\ell_j^2} \frac{C\,\ell_j\sigma}{1+\ell_j\sigma}\\
\geq N_{s,\omega}\cdot \frac{C\,\sigma}{\tilde\ell(1+\tilde\ell\sigma)}=N_{s,\omega}\cdot \mu\, \frac{C}{s^2}\,\frac{\sigma\tilde\ell}{1+\sigma\tilde\ell}.
\end{multline}
 For the sum over the large intervals we use \eqref{tos} to estimate
\begin{multline}\label{3.7}
\sum_{\ell_j\geq \tilde \ell}\frac {n_j^{\rm GP}}{\ell_j^2}e(n_j^{\rm GP}\ell_j\gamma, \ell_j\sigma)\geq
\inf_{\sum n_i=1-N_{s,\omega}}\sum_{\ell_j\geq \tilde \ell}\frac{n_j}{\ell_j^2}e(n_j\ell_j\gamma,\ell_j\sigma)\\ \geq 
\inf_{\sum n_i=1-N_{s,\omega}}\sum_{j}\frac{n_j}{\ell_j^2}e(n_j\ell_j\gamma,\infty)(1-K(\tilde \ell \sigma)^{-1/2}).
\end{multline}
Apart from the factor $(1-K(\tilde \ell \sigma)^{-1/2})$ the right side is the GP energy for $\sigma=\infty$ with normalization $\int|\psi|^2=1-N_{s,\omega}$ instead of $\int|\psi|^2=1$. By simple scaling this is $1-N_{s,\omega}$ times the the GP energy
with normalization 1 and $\gamma$ replaced by $(1-N_{s,\omega})\gamma$, which in turn is not smaller than $(1-N_{s,\omega})^2$ times $e^{\rm GP}_\omega(\gamma,\nu,\infty)$. We can further estimate 
\beq \label{3.8}(1-N_{s,\omega})^2e^{\rm GP}_\omega(\gamma,\nu,\infty)\geq (1-2N_{s,\omega})e^{\rm GP}_\omega(\gamma,\nu,\sigma)\eeq 
and putting  (\ref{3.6}), (\ref{3.7}) and (\ref{3.8}) together we obtain
\beq\label{3.9}
\frac{e^{\rm GP}_\omega(\gamma,\nu,\sigma)}\mu\geq N_{s,\omega}\cdot \frac{C}{s^2}\,\frac{\sigma\tilde\ell}{1+\sigma\tilde\ell}
+(1-2N_{s,\omega})\frac{e^{\rm GP}_\omega(\gamma,\nu,\sigma)}\mu(1-K(\tilde \ell \sigma)^{-1/2}).
\eeq
If  $\nu$, $\gamma$ and $\sigma$ tend to infinity under the constraints \eqref{conditions},  the ratio ${e^{\rm GP}_\omega(\gamma,\nu,\sigma)}/\mu$ stays bounded (in probability) according to Theorem 3.1 in \cite{SYZ}. Moreover, for
if $\gamma\lesssim \nu^2$ we have by (\ref{3.3}) and \eqref{conditions}
\beq\label{sigmaell}
\sigma \tilde \ell=s\sigma/\sqrt \mu \gg 1.
\eeq
For $C/s^2>2 e^{\rm GP}_\omega(\gamma,\nu,\sigma)/\mu$ we thus arrive at an estimate for the mass in the small intervals:
\beq\label{3.10}
N_{s,\omega}\leq\text{(const.)}\frac{e^{\rm GP}_\omega(\gamma,\nu,\sigma)}\mu\cdot \frac{\mu^{1/4}}{\sigma^{1/2}},
\eeq
and since, by \eqref{sigmaell},
\beq\label{3.11} {\mu^{1/4}}/{\sigma^{1/2}}\ll 1
\eeq
we have shown that $N_{s,\omega}\to 0$ in probability if $\gamma\lesssim \nu^2$ and the conditions \eqref{conditions} holds.

Now according to \eqref{superfbound} the superfluid fraction is bounded from above by $N_{s,\omega}/L_\omega^2$ where $L_\omega$ is the total length of intervals of length $\leq\tilde \ell$. The latter converges in probability to the expectation value
\beq  L=\nu\int_0^{\tilde\ell}\ell dP_\nu(\ell)=1-(1+(\nu\tilde\ell))e^{-\nu\tilde\ell}),\eeq
provided the fluctuations remain small.  
For $\gamma \ll \nu^2$ we have $\tilde\ell \nu\gg 1$ (by \eqref{3.3}) and the length $L$ converges to 1 as $\nu\to \infty$, while for $\gamma \sim \nu^2$ the length stays bounded away from 0 because $\tilde\ell \nu$ is $O(1)$. The fluctuations are $O(\nu^{-1/2})$. Hence the superfluid fraction tends to 0 in probability for $\gamma\lesssim \nu^2$.

\subsection{The case $\gamma\gg \nu^2$} Here $\mu\sim\gamma$ and we take $\tilde\ell\sim \mu^{-1/2}\sim \gamma^{-1/2}\ll\nu^{-1}$.
We need in any case $\sigma\tilde \ell \gg1$, i.e., $\sigma\gg \gamma^{1/2}$, which is compatible with the conditions \eqref{conditions}. In the same way as above we obtain \eqref{3.10}, this time with $\mu\sim\gamma$.

Since $\nu\tilde \ell\sim \nu/\gamma^{1/2}\ll 1$, however, the average length of the small intervals is now $L\sim (\nu/\gamma^{1/2})^2\ll1$ rather than $O(1)$ as for $\gamma\lesssim \nu^2$. To exclude superfluidity we need
\beq\label{sfest} N_{s,\omega}/L^2\sim(\gamma^{1/4}/\sigma^{1/2})(\gamma/\nu^2)\ll 1\eeq
which holds for \beq\label{sigmagg} \sigma\gg (\gamma/\nu^2)^4\gamma^{1/2}.\eeq
This condition is still not sufficient, however, because the estimate $L_\omega\sim (\nu/\gamma^{1/2})^2$ can only be claimed to be true in probability as long as the fluctuations of the random variable $L_\omega=\sum_{\ell_j\leq\tilde\ell}\ell_j$ are small compared to its average value, $L$.  A sufficient condition for this is that $\nu\int_0^{\tilde\ell}\ell^2dP_\nu(\ell)\ll L^2$, which holds for $\gamma\ll \nu^4$. Altogether we conclude that the superfluid fraction tends to 0 in probability, if \eqref{sigmagg} together with $\nu^2\ll\gamma\ll \nu^4$ hold.

\section{Concluding remarks}

We have studied superfluidity in the ground state of a one-dimensional model of bosons with a repulsive contact interaction and in a random potential generated by Poisson distributed point obstacles. In the Gross Pitaevskii (GP) limit this model always shows complete BEC, but depending on the parameters, superfluidity may or may not occur. In the course of the analysis we derived the closed formula \eqref{defsf} for the superfluid fraction, expressed in terms of the GP wave function. 

The advantage of this model is that it is amenable to a rigorous mathematical analysis leading to unambiguous statements.  It has its limitations: Nothing is claimed about positive temperatures and the proof of BEC requires that the ratio between the coupling constant for the interaction and the density tends to zero as $N\to\infty$. Nevertheless, to our knowledge this is the only model where a Bose glass phase in the sense of complete BEC but absence of superfluidity (cf.\ \cite{YY}) has been rigorously established so far.

\medskip
\section*{Acknowledgements}
This work is supported by the Austrian Science Fund (FWF) under project P 22929-N16.

\end{document}